\documentstyle[aps,pre,psfig]{revtex}
\begin{document}
\draft
\twocolumn[\hsize\textwidth\columnwidth\hsize\csname@twocolumnfalse\endcsname
\title{Universal behavior of localization of residue fluctuations in
globular proteins}
\author{Yinhao Wu, Xianzhang Yuan, Xia Gao, Haiping Fang, and Jian Zi\cite%
{byline}}
\address{Surface Physics Laboratory (National Key Lab) and\\
T-Center for Life Sciences, Fudan University, Shanghai 200433, People's
Republic of China}
\date{\today}
\maketitle

\begin{abstract}
Localization properties of residue fluctuations in globular proteins are
studied theoretically by using the Gaussian network model. Participation
ratio for each residue fluctuation mode is calculated. It is found that the
relationship between participation ratio and frequency is similar for all
globular proteins, indicating a universal behavior in spite of their
different size, shape, and architecture.
\end{abstract}


]\narrowtext

Proteins are important biological macromolecules that control almost all
functions of living organisms. It was once believed that proteins are rather
amorphous and without well-defined structures. After more and more
structures have been determined by crystallographic and NMR methods, it has
revealed that protein structures are far from random. They have well-defined
secondary and tertiary structures which comprise essential information
relating to their functions and mechanisms.

Proteins in the folded states are not static. Instead, the constituent
residues fluctuate near their native positions owing to the finite
temperature effects\cite{fra:91}. It has been now well accepted that the
fluctuations are crucial for enzyme catalysis and for biological activity%
\cite{sto:99,zac:00}. Recently, there has been considerable interest in the
correlations between protein functions and fluctuations\cite{sto:99}.
Intensive theoretical studies on fluctuations of protein have been carried
out based on either molecular dynamics simulations or normal mode analyses
(NMA) by using all-atom empirical potentials\cite{kit:99}. It has been shown
that the NMA is a very useful method to study protein fluctuations\cite%
{lev:85,hay:97}. The use of atomic approaches becomes computational
demanding when dealing with large proteins. For proteins composed of more
than thousand residues, it is difficult to investigate by using the
conventional atomic models and potentials. On the other hand, coarse-grained
protein models and simplified force fields have revealed a great success in
description of the residue fluctuations of proteins\cite%
{tir:96,hal:97,bah:98,hin:99}. Although there have been intensive studies on
residue fluctuations, to our knowledge, there is few study on localization
properties of residue fluctuations.

In this paper, based on a coarse-grained protein model, we show
theoretically that there is a similar behavior in the localization of
residue fluctuations for globular proteins, even though their architectures
and sizes are rather different. In our study of residue fluctuations,
proteins are modeled as elastic networks. The nodes are residues linked by
inter-residue potentials that stabilizes the folded conformation. This model
has been usually referred as the Gaussian network model (GNM), which can
give a satisfactory description of the fluctuation of folded proteins\cite%
{hal:97,bah:98,bah:97,dem:98,jer:99,kes:00,ati:01}. In this model, residues
are assumed to undergo Gaussian-distributed fluctuations about their native
positions. No distinction is made between different types of residues. A
single generic harmonic force constant is used for the inter-residue
interaction potential within a cutoff range. We consider residues as the
minimal representative units and the $\alpha $-carbons are used as
corresponding sites for residues. Considering all contacting residues, the
internal Hamiltonian within the GNM is given by\cite{hal:97,bah:98} 
\begin{equation}
H=\frac{1}{2}\gamma \left[ \Delta {\bf R}^{\text{T}}\left( {\bf \Gamma }%
\otimes {\bf E}\right) \Delta {\bf R}\right] ,
\end{equation}%
where $\gamma $ is the harmonic force constant; $\left\{ \Delta {\bf R}%
\right\} $ represents the $3N$-dimensional column vectors of fluctuations $%
\Delta {\bf R}_{1},\ldots ,\Delta {\bf R}_{N}$ of the $C^{\alpha }$ atoms,
where $N$ is the number of residues; ${\bf E}$ is the third order identity
matrix; the superscript T denotes the transpose; $\otimes $ stands for the
direct product, and ${\bf \Gamma }$ is the $N\times N$ Kirchhoff matrix\cite%
{har:71} with the elements given by 
\begin{equation}
\Gamma _{ij}=\left\{ 
\begin{array}{ll}
-H\left( r_{\text{c}}-r_{ij}\right) , & i\neq j, \\ 
-\sum\limits_{i(\neq j)}^{N}\Gamma _{ij}, & i=j.%
\end{array}%
\right. 
\end{equation}%
Here, $r_{ij}$ is the separation between the $i$-th and $j$-th $C^{\alpha }$
atoms; $H(x)$ is the Heaviside step function, and $r_{\text{c}}$ is the
cutoff distance outside of which there is no inter-residue interaction. The $%
i$-th diagonal element of $\Gamma $ characterizes the local packing density
or the coordination number of residue $i$. The inverse of the Kirchhoff
matrix can be decomposed as 
\begin{equation}
\Gamma ^{-1}={\bf U}\Lambda ^{-1}{\bf U}^{\text{T}},  \label{gamma}
\end{equation}%
where ${\bf U}$ is an orthogonal matrix whose columns ${\bf u}_{i}\left(
1\leq i\leq N\right) $ are the eigenvectors of $\Gamma $, and $\Lambda $ is
diagonal matrix of eigenvalue $\lambda _{i}$ of $\Gamma $.
Cross-correlations of residue fluctuations between the $i$-th and $j$-th
residues are found from 
\begin{equation}
\left[ \Delta {\bf R}_{i}\cdot \Delta {\bf R}_{j}\right] =\frac{3k_{B}T}{%
\gamma }\left[ \Gamma ^{-1}\right] _{ij}.  \label{corr}
\end{equation}%
From Eqs. (\ref{gamma}) and (\ref{corr}), the mean-square (ms) fluctuations
(also called Debye-Waller or B-factors) of the $i$-th residue associated
with the $\alpha $-th mode are given by 
\begin{equation}
\left[ \Delta {\bf R}_{i}\cdot \Delta {\bf R}_{i}\right] _{\alpha }=\frac{%
3k_{B}T}{\gamma }\lambda _{\alpha }^{-1}\left[ u_{\alpha }\right] _{i}\left[
u_{\alpha }\right] _{i}.
\end{equation}

In our calculation, the cutoff distance $r_{\text{c}}=7$ \AA\ is used, as
adopted in previous studies\cite{hal:97,bah:98}. The harmonic force constant 
$\gamma $ is determined by fitting to the experimental ms fluctuations. From
this model, one can obtain the fluctuation mode frequencies and eigenvectors
for a given protein. The GNM can in general give results in good agreement
with the observed B-factors \cite{hal:97,bah:98}.

The spatial distribution of a given mode is characterized by its
eigenvectors. To study the localization properties of protein fluctuations,
we have to compute participation ratio (PR) for each mode, defined by\cite%
{bel:72} 
\begin{equation}
P_{\alpha }=\frac{1}{N}\left( \sum\limits_{i}\left[ u_{\alpha }\right]
_{i}^{4}\right) ^{-1}.
\end{equation}%
Values of PR range from $1/N$ to unity. PR takes the value of unity if all
residues have equal fluctuation. If only one residue fluctuates PR is equal
to $1/N$. From its definition, it is obvious that PR is a measure of the
degree of localization. If the PR is small for a given mode, only a few
residues have considerable fluctuations and the mode is a localized one. On
the other hand, if the PR is large for a given mode, the mode is delocalized.

It is known that at the physiological temperatures, protein fluctuates among
different conformations around its native one. Therefore, in principle, all
contributions from these conformations should be considered in the
calculation of PR. Unfortunately, only one conformation could be obtained
from experiments. However, these conformations could be obtained
approximately by the following way. For each residue, it is assumed that it
can stay at any position inside the sphere with a radius of half the
magnitude of fluctuation centered on the position obtained from experiments.
A conformation can be derived by a random choice of the position for each
residue while the inter-distance between two adjacent residues is kept
unchanged within the framework of the SHAKE algorithm \cite{ryc:77}.

\begin{figure}[tbp]
\centerline{\psfig{figure=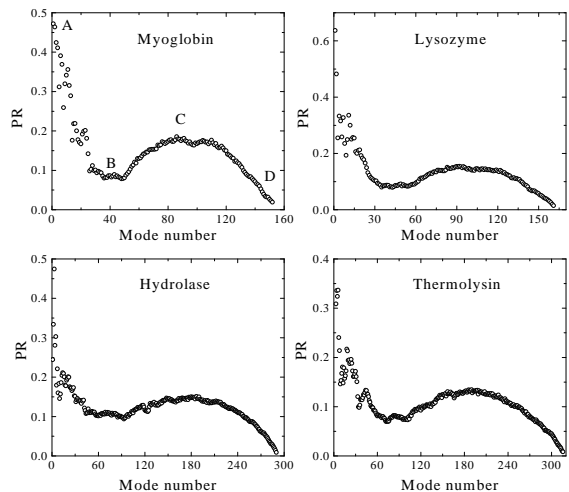,width=3in,clip=}}
\caption{Calculated participation ratio of residue fluctuations for proteins
listed in Table \protect\ref{tab1}.}
\label{fig1}
\end{figure}

The calculated PR for several proteins is shown in Fig. \ref{fig1}. The
Brookhaven Protein Databank (PDB) codes and references of the proteins
studied are listed in Table \ref{tab1}. The modes are numbered starting from
the lowest frequency. In the calculations about 100 conformations are
adopted. It is found that if more conformations are used, the curves will
become smoother eventually.

Based on the Anderson localization theory\cite{and:58,and:78}, Bahar {\it et
al}. \cite{bah:98} suggested that modes with larger fluctuation frequencies
would be more localized, indicating a monotonous decrease in PR with
frequency. As suggested by Onuchic {\it et al}. \cite{onu:00} proteins are
neither ordered nor random systems, the localization properties of protein
fluctuations should show some intrinsic features from those in ordered or
random systems.

It can be seen from Fig. \ref{fig1} that starting from lowest frequency, PR
first decreases with frequency, then increase, and finally decreases with
frequency. A large number of globular proteins which have diversified
topology, secondary structure arrangement and size are calculated. This
behavior of PR seems to be universal, holding for all globular proteins.
Other molecular systems such as tRNA are also calculated. But the behavior
of PR is qualitatively different from proteins (data not shown). So it is
reasonable to conjecture that the different behavior of PR in globular
proteins from other systems reflects the intrinsic difference of certain
properties. Recently, Micheletti {\it et al. }\cite{mic:02} studied the
localization properties of HIV-1 protease. A similar behavior of PR in HIV-1
protease was found.  

\begin{figure}[tbp]
\centerline{\psfig{figure=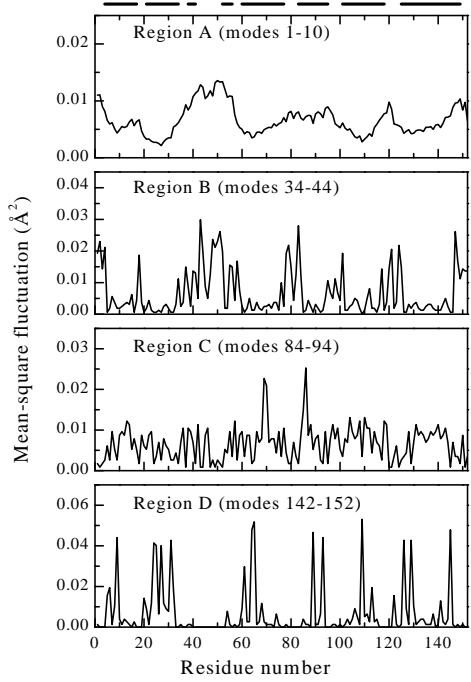,width=3in,clip=}}
\caption{Calculated projected residue ms fluctuations in different frequency
regions for myoglobin. The secondary structures of myoglobin are represented
by the horizontal segment heavy lines at the top of the figure. The
remaining are loops.}
\label{fig2}
\end{figure}

To study the origin of the behavior of PR in globular proteins, the
fluctuation patterns of the protein {\it myoglobin} at different frequency
regions are given in Fig. \ref{fig2}. The different frequency regions in the
figure are labeled by different letters (see Fig. \ref{fig1}). In the low
frequency region A, the fluctuations represent a collective motion,
characterized by large values of PR. In the region B, the PR is small,
implying localized fluctuations. It is interesting to note that in this
region the fluctuations occur dominantly at the loops. In the highest
frequency region D, the fluctuations are found to be confined to the
secondary structures, resulting small PR. In the region C, one can find that
motions of both loops and secondary structures are involved. The degree of
localization is, however, smaller than that in regions B and D, but it is
larger than that in region A. Therefore, it can be concluded from the
fluctuation patterns that the dip of PR occurred at lower frequency side
(region B) originates from the localized fluctuations at loops that connect
the secondary structures. For conventional disordered solids or random
coils, there are nearly no well-defined secondary structures and
consequently no loops. The resulting PR will show a different behavior. It
is obvious that the different behavior of PR in globular proteins from that
of conventional random solids or coils originates from the different nature
of structures.

\begin{figure}[tbp]
\centerline{\psfig{figure=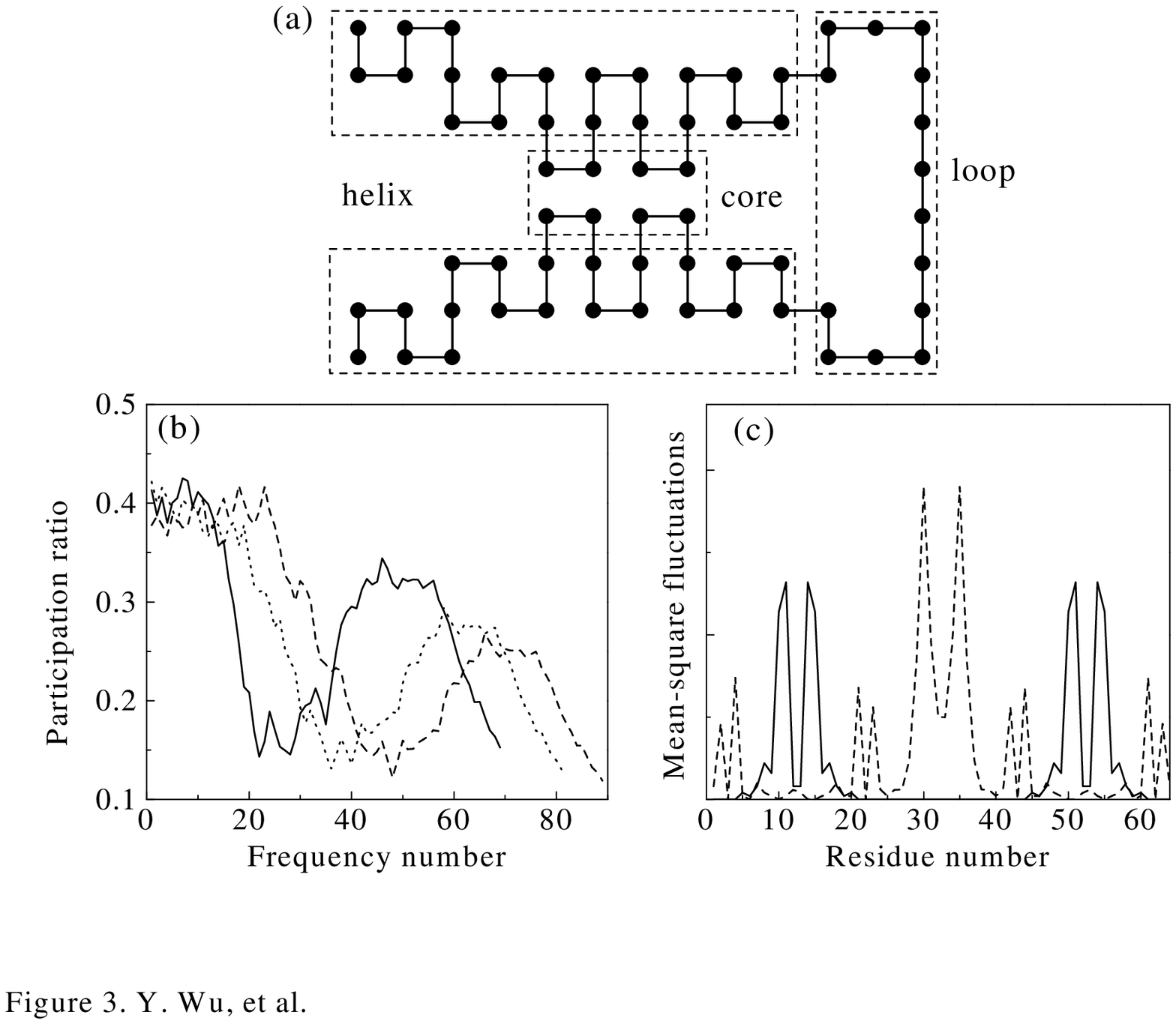,width=3in,clip=}}
\caption{(a) Lattice model protein consists of two helices, a loop and a
core region. The loop length can be changed by moving the loop horizontally
to the right. (b) Participation ratio for model proteins with different loop
length. Solid line is for the model protein shown in (a) with loop length of
13$a$, where $a$ is the lattice constant. Dotted and dashed lines are for
model proteins with loop length of 23$a$ and 33$a$, respectively. (c)
Projected residue ms fluctuations in arb. units for the modes with smallest
PR in the dip region (dashed line) and with largest frequency (solid line).}
\label{fig3}
\end{figure}

To get a deeper insight into how the localization properties are affected by
the topology, a lattice model\cite{dil:95} with different length of the loop
is adopted. In this model, a protein is represented by a self-avoiding chain
of beads placed on a two-dimensional discrete lattice. In construction of
this model protein, one must consider the fact that the secondary structure
has higher packing density while the loop has lower packing density. A core
region is introduced by making two helices contacted each other since cores,
with higher packing density, are important to stabilize the whole structure.
Our model protein shown in Fig. \ref{fig3}(a) consists of two helices, a
connective loop, and a core. All residues (beads) are treated identically.
In our calculations only the nearest neighbor interaction is considered.

Advantages of the lattice model are that we can change the structure as
desired to get insight into how the residue fluctuations are affected by the
changes in structures, which is difficult to do in real proteins. In Fig. %
\ref{fig3}(b) the calculated PR by the GNM for the model protein with
different loop length is shown. The loop length is changed by moving the
loop horizontally to the left or right. The curves are smoothed simply by
adjacent averaging using 10 points. It is obvious that the PR of
fluctuations in the simple model protein shows a similar behavior to that of
real globular proteins. With increase in the loop length, the PR values of
both the dip (region B in Fig. \ref{fig1}) and the peak (region C in Fig. %
\ref{fig1}) decrease. It can be seen from Fig. \ref{fig3}(c) that at the dip
the fluctuations are dominant in the loop region. Again, the origin of the
dip is the cause of the loop. For the highest frequency mode, the
fluctuations dominantly occur at the helices, especially at the core region.
The broad peak comprises modes which are more delocalized and worse defined.
These peaks are relevant to the coupling motions among secondary structures.

In summary, localization properties of fluctuations in globular proteins
were studied by using the Gaussian network model. It was found that the
participation ratio of fluctuations in globular proteins shows a universal
behavior, confirmed by theoretical calculations in both real globular and
model proteins. The loops connecting the secondary structures are
responsible for this feature.

This work was supported by the NSFC. Partial support from Shanghai Science
and Technology Commission, China is acknowledged. Interesting discussions
with Dr. Y. Q. Zhou, Dr. C. Tang, and Dr. J. Z. Y. Chen are acknowledged.

\begin{table}[tbp]
\caption{The PDB code and reference of proteins studied in the present work.}
\label{tab1}%
\begin{tabular}{ccc}
Protein & PDB code & Reference \\ 
\tableline 
\begin{tabular}{l}
Myoglobin \\ 
Lysozyme \\ 
Hydrolase \\ 
Thermolysin%
\end{tabular}
& 
\begin{tabular}{l}
1bvc \\ 
166l \\ 
1amp \\ 
5tln%
\end{tabular}
& 
\begin{tabular}{l}
\onlinecite{vag:95} \\ 
\onlinecite{wea:87} \\ 
\onlinecite{sch:92} \\ 
\onlinecite{hol:81}%
\end{tabular}%
\end{tabular}%
\end{table}


\begin{references}
\bibitem[*]{byline} To whom all correspondence should be addressed.
Electronic address: jzi@fudan.edu.cn

\bibitem{fra:91} H. Frauenfelder, S. G. Sligar, and P. G. Wolynes, Science 
{\bf 254}, 1598 (1991).

\bibitem{sto:99} A. Stock, Nature {\bf 400}, 221 (1999).

\bibitem{zac:00} G. Zaccai, Science {\bf 288}, 1604 (2000).

\bibitem{kit:99} A. Kitao and N. Go, Curr. Opin. Struct. Biol. {\bf 9}, 164
(1999) and references therein.

\bibitem{lev:85} M. Levitt, C. Sander, and P. S. Stern, J. Mol. Biol. {\bf %
181}, 423 (1985).

\bibitem{hay:97} S. Hayward, A. Kitao, and H. J. C. Berendsen, Proteins {\bf %
27}, 425 (1997).

\bibitem{tir:96} M. M. Tirion, Phys. Rev. Lett. {\bf 77}, 1905 (1996).

\bibitem{hal:97} T. Haliloglu, I. Bahar, and B. Erman, Phys. Rev. Lett. {\bf %
79}, 3090 (1997).

\bibitem{bah:98} I. Bahar {\it et al}., Phys. Rev. Lett. {\bf 80}, 2733
(1998).

\bibitem{hin:99} K. Hinsen and G. R. Kneller, J. Chem. Phys. {\bf 111},
10766 (1999).

\bibitem{bah:97} I. Bahar and R. L. Jernigan, J. Mol. Biol. {\bf 266}, 195
(1997); I. Bahar, A. R. Atilgan, and B. Erman, Fold. Des. {\bf 2}, 173
(1997); I. Bahar and R. L. Jernigan, J. Mol. Biol. {\bf 281}, 871 (1998); I.
Bahar {\it et al}., Biochem. {\bf 37}, 1067 (1998); I. Bahar {\it et al}.,
J. Mol. Biol. {\bf 285}, 1023 (1999); I. Bahar and R. L. Jernigan, Biochem. 
{\bf 38}, 3478 (1999).

\bibitem{dem:98} M. C. Demirel {\it et al}., Protein Sci. {\bf 7}, 2522
(1998).

\bibitem{jer:99} R. L. Jernigan, M. C. Demirel, and I. Bahar, Int. J. Quant.
Chem. {\bf 75}, 301 (1999).

\bibitem{kes:00} O. Keskin, I. Bahar, and R. L. Jernigan, Biophys. J. {\bf 78%
}, 2093 (2000).

\bibitem{ati:01} A. R. Atilgan {\it et al}., Biophys. J. {\bf 80}, 505
(2001).

\bibitem{har:71} F. Harry, {\it Graph Theory} (Addison-Wesley, Reading, MA,
1971).

\bibitem{bel:72} R. J. Bell, Rep. Prog. Phys. {\bf 35}, 1315 (1972).

\bibitem{ryc:77} J. P. Ryckaert, G. Ciccotti G, H. J. C. Berendsen, J.
Comput. Phys. {\bf 23}, 327 (1977).

\bibitem{vag:95} U. G. Wagner {\it et al.}, J. Mol. Biol. {\bf 247}, 326
(1995).

\bibitem{wea:87} L. H. Weaver and B. W. Matthews, J. Mol. Biol. {\bf 193},
189 (1987).

\bibitem{sch:92} C. Schalk {\it et al.}, Arch. Biochem. Biophys. {\bf 294},
91 (1992).

\bibitem{hol:81} M. A. Holmes and B. W. Matthews, Biochem. {\bf 20}, 6912
(1981).

\bibitem{and:58} P. W. Anderson, Phys. Rev. {\bf 109}, 1492 (1958).

\bibitem{and:78} P. W. Anderson, Rev. Mod. Phys. {\bf 50}, 191 (1978).

\bibitem{onu:00} J. N. Onuchic {\it et al}., Adv. Protein Chem. {\bf 53}, 87
(2000).

\bibitem{dil:95} K. A. Dill {\it et al.}, Protein Sci. {\bf 4}, 561 (1995).

\bibitem{mic:02} C. Micheletti, G. Lattanzi, and A. Maritan, J. Mol. Biol. 
{\bf 321}, 909 (2002).
\end{references}
\end{document}